\documentclass[journal=jacsat,manuscript=article,layout=twocolumn]{achemso}

\usepackage[version=3]{mhchem} 
\usepackage[T1]{fontenc}       

\usepackage{xcolor}

\usepackage[colorinlistoftodos]{todonotes}

\author{Novoselov D.}
\affiliation{Institute of Metal Physics, S.Kovalevskoy St. 18, 620990 Yekaterinburg, Russia}
\alsoaffiliation{Ural Federal University, Mira St. 19, 620002 Yekaterinburg, Russia}
\email{novoselov@imp.uran.ru}

\author{Korotin Dm.M.}
\affiliation{Institute of Metal Physics, S.Kovalevskoy St. 18, 620990 Yekaterinburg, Russia}

\author{Anisimov V.I.}
\affiliation{Institute of Metal Physics, S.Kovalevskoy St. 18, 620990 Yekaterinburg, Russia}
\alsoaffiliation{Ural Federal University, Mira St. 19, 620002 Yekaterinburg, Russia}

\title[Quantum states entanglement in hemoglobin molecule active center]
{Quantum states entanglement in hemoglobin molecule active center}

\abbreviations{DFT, DFT+DMFT}

\begin{document}







\begin{abstract}
An {\em ab initio} study of the electronic and spin configuration for the iron ion in the active center of the human hemoglobin molecule is presented. 
It is well known that the iron ion, being surrounded by the porphyrin ring and the ligands, plays the key role in the realization of the basic oxygen-transport functions of the molecule.
This work is focused on the investigation the features of the 3$d$-shell electronic states of the iron ion located inside the active center of the hemoglobin molecule.
Also in this paper we study in detail the changes in these states occurring during the oxidation process.
We use a combination of the Density Functional Theory (DFT) method and the Dynamical Mean Field Theory (DMFT) approach.
This method allows to consider dynamic correlation effects that are important in the description of systems containing transition metal ions.
It was found that the state of the valence electrons of the iron ion of the active center of hemoglobin molecule is the entangled quantum state.
This state is a mixture of several electronic states with comparable statistical probability.
Furthermore, it was found that the process of the bond formation between the iron-porphyrin complex and the oxygen molecule is more complex than a simple high-spin to low-spin state of the Fe ion transition. The transition metal ion oxidation is accompanied by substantial redistribution of the states probabilities and the increasing of the entanglement degree.
This process also leads to the reduction of the total spin moment from $s\approx$2.1 for the FeP(Im) to $s\approx$1.7 for the FeP(Im)(O$_2$).
\end{abstract}


\section{Introduction}

Metal-porphyrin complexes play an important role in many biological processes such as photosynthesis process with a chlorophyll assistance in plant cells or the transport of oxygen by hemoglobin in living beings through reversible binding of a hemoglobin molecule with an oxygen molecule as well as they widely used for therapeutic purposes~\cite{Chandra2000}.
It is not surprising that the interest to this kind of system isn't being quenched until now~\cite{PhysRevB.79.245404,PhysRevB.82.081102,Radon2008a}.
Undoubtedly, one of the most interesting features of the hemoglobin molecule is an ability to reversibly attach itself to oxygen molecules. 
This process is accompanied by the spin state transition of the iron ion. 
A number of papers~\cite{Hu2005,Panchmatia2008b,Scherlis2007a,Panchmatia2010a} indicate that the transition occurs between closely energy located low- and high-spin states. 
In addition it is noted that there are energy close-lying excited states with unknown spin multiplicity~\cite{Hu2005}.
At the same time, in the works~\cite{Radon2008a,Weber2013} the assumption of the mixed nature of the spin state is discussed, pointing to the possibility of existence a contributions of several atomic states with different values of valence and spin moment. 
This indicates that the detailed and complete interpretation of the spin transition scenario in the hemoglobin molecule during binding to the oxygen molecule is currently insufficient.
Thus a detailed study of quantum state nature of the iron ion in the hemoglobin molecule and evaluation the states during oxidation is undoubtedly interesting.

In the works~\cite{Scherlis2007a,Panchmatia2008b,Panchmatia2010a,Weber2013} it was shown that the electronic correlations play an important role in the iron-porphyrin complexes. Therefore, we have selected a modern {\em ab initio} approach that combines the density functional theory and the dynamical mean field theory (DFT+DMFT) to describe the states of the iron ion.
This method allows one to take into account many-body effects in paramagnetic systems as well as thermal fluctuation effects. Through applying the continuous time quantum Monte-Carlo algorithm (CT-QMC) as an impurity solver it is possible to perform a statistical analysis of the atomic states of the impurity site.

\section{Method and structural model}
\begin{figure}[ht]
\centerline{\includegraphics[width=0.8\linewidth]{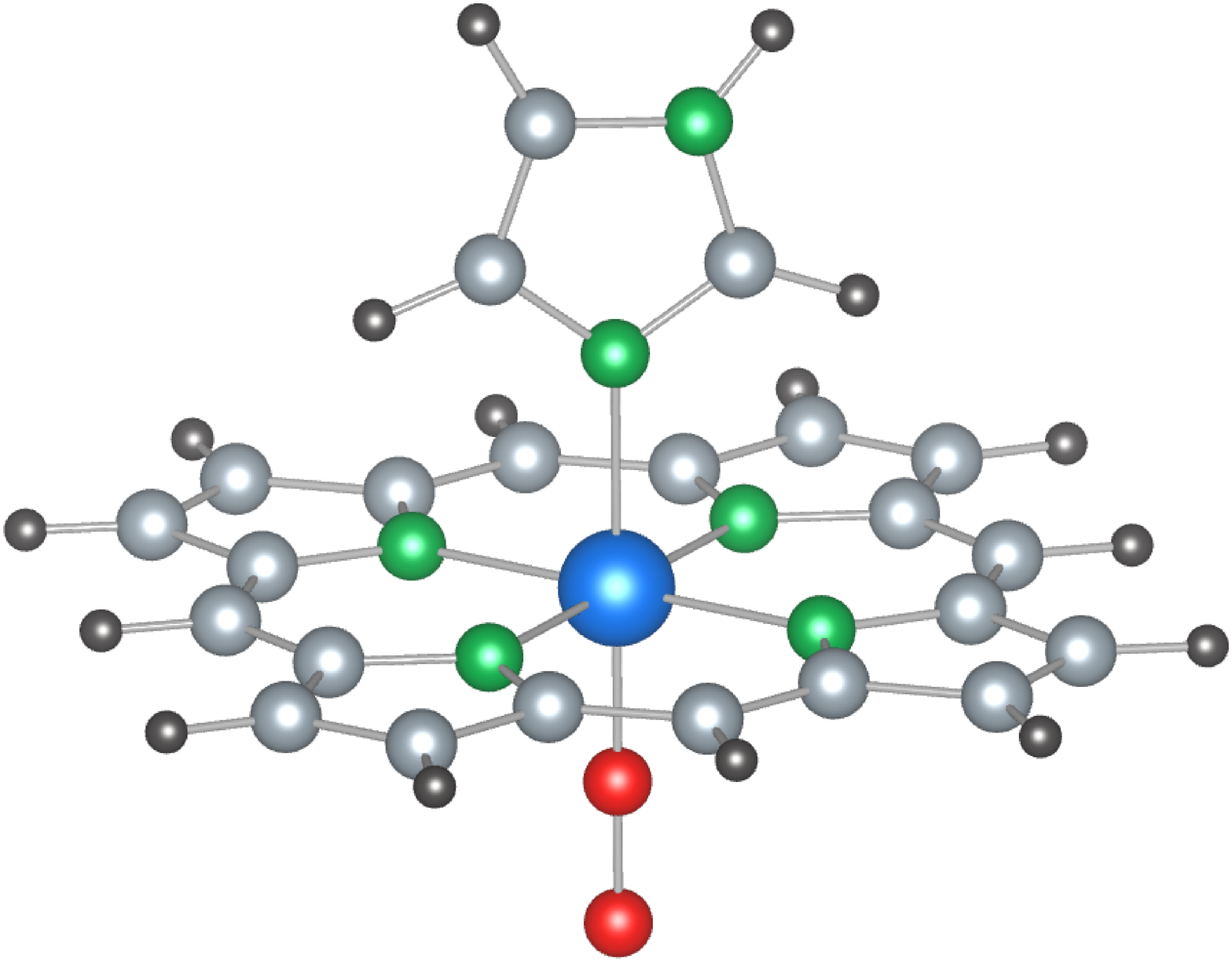}}
\caption{The schematic image of structure model of active center of the human hemoglobin molecule. The blue sphere corresponds to the iron atom, the red spheres -- the oxygen atoms, the dark grey -- the hydrogen atoms, the light grey -- the carbon atoms, green -- the nitrogen atoms.}
\label{fig:structure}
\end{figure}

The human hemoglobin molecule consist of several thousand atoms. 
In this work the model is investigated (see Figure \ref{fig:structure}) including only the active center of the molecule that directly involved in the formation of the chemical bond with the oxygen molecule.

This structural model is based on the existing experimental data~\cite{Park2006,Fermi1984,Shaanan1983}.
It includes the ferroprotoporphyrin and two ligands - proximal and distal, playing a leading role in the functional activity of the hemoglobin molecule.
The active center is formed by almost flat porphyrin ring consisting of nitrogen and carbon atoms. 
At the center of the ring the iron ion is placed. 
The nearest neighbours of the transition metal ion are four nitrogen atoms of the porphyrin ring. Fifth nearest neighbour of the iron ion is the nitrogen from the imidazole (Im) molecule that is located almost perpendicular to the plane of the porphyrin-metal complex (see Figure~\ref{fig:structure}). 
Therefore the transition metal ion is surrounded by the pyramid of the nitrogen atoms.

In this paper the structural model containing 46 atoms for the deoxy-molecule and 48 atoms for the oxy-HHB was considered. We started from the hemo group of the d-chain of real HHB molecule~\cite{Fermi1984}. Then the hemo group was enforced to be flat and located in a $xy$-plane, the iron ion was placed in the center of the hemo ring and the missing chemical bonds were saturated by the hydrogen atoms. Then an atomic relaxation for the obtained FeP crystal structure was performed. 

The calculations were done with Quantum-ESPRESSO package~\cite{Giannozzi2009} within the ultrasoft pseudopotential formalism in plane waves basis. We used GGA (PBE) exchange correlation functional, plane waves energy cutoff was set to 75~Ry. The modeled molecule was placed in the center of the cubic cell with the edge length of 40 a.e.

To obtain the FeP(Im) structure, we placed the imidazol molecule at 2.13~\AA~\cite{Fermi1984} distance from the iron ion of the FeP group and the imidazole molecule atoms positions were relaxed keeping Fe-N(Im) distance constant.

To get the last structure -- FeP(Im)(O$_2$), the dioxygen molecule was placed on the opposite to the Im molecule side of the FeP plane at 1.76~\AA~\cite{Shaanan1983} distance and then relaxed keeping the Fe-O distance fixed.

As a result of this procedure we have obtained two relaxed molecular structures that are simple and symmetric enough to analyze the splitting of the energy levels of the Fe ion placed in the center. On the other hand, local surroundings of the Fe ion remains close to the experimental data.
The distances Fe-Im and Fe-O$_2$ are kept fixed to 2.13~\AA ~ and 1.76~\AA ~ respectively~\cite{Fermi1984,Shaanan1983}. 
The obtained distance between the Fe and N atoms of the porphyrin ring equals 2.07~\AA, that is close to the experimental value 2.05~\AA~\cite{Fermi1984}.

To take into account electronic correlations in the $d$-shell of the iron ion the DMFT approach 
was used. As an input this approach requires an uncorrelated Hamiltonian matrix, typically obtained from the DFT calculation, and the values of Coulomb interaction parameter and Hund exchange parameter. To obtain the Hamiltonian matrix we used the projection procedure to the basis of Wannier functions~\cite{Korotin2008}. All energy bands and all atomic states for all atoms of the system were used for projection. 
Therefore, the Wannier functions coincide with atomic orbital by the construction. The five Wannier functions with symmetry of the five Fe-$d$ atomic orbitals were considered as correlated wavefunctions. When we refer to a some $d$-state occupation below, for example the $x^2-y^2$ state, we mean the occupation number of the Wannier function centered on the Fe ion with the corresponding spacial symmetry.

It is stated in the work~\cite{Weber2013} that the results of the DMFT calculation for HHB molecule show the weak dependence on the temperature used in the calculation. We carried out the DMFT calculations with the inverse temperature parameter $\beta$=25.
For the construction of the interaction matrix the parameter of Coulomb repulsion ($U$=4~eV) and Hund exchange parameter ($J$=0.9~eV) were chosen to be the same as in earlier works~\cite{Weber2013,PhysRevB.82.081102,Scherlis2007a}.

Subsequently the obtained Hamiltonian was solved within the DMFT approach using the AMULET package~\cite{amulet}. 
DMFT is applicable to a wide class of physical systems with the presence of strong correlation effects.
This method reduces a many body lattice problem to the impurity one~\cite{Metzner1989,Anisimov1997} and it considers the single atom placed in the external environment with allowed the electrons swap between the impurity and the environment. 
To solve the auxiliary impurity problem we used continuous time quantum Monte-Carlo algorithm (CT-QMC)~\cite{RevModPhys.83.349,Werner2006a,Werner2006}.
The hybridization expansion CT-QMC solver provides the site-reduced statistical operator (density matrix)~\cite{PhysRevLett.99.126405}. 
This quantity describes the probability of finding an atom in a particular many-body state and an expectation value of any local operator can be easily obtained from it. Therefore, this instrument is well suited to analyze a statistical probability of the various atomic configurations of the Fe ion in the HHB active center.

\section{Results and discussion}

\begin{figure}[ht]
\centerline{\includegraphics[width=0.9\linewidth]{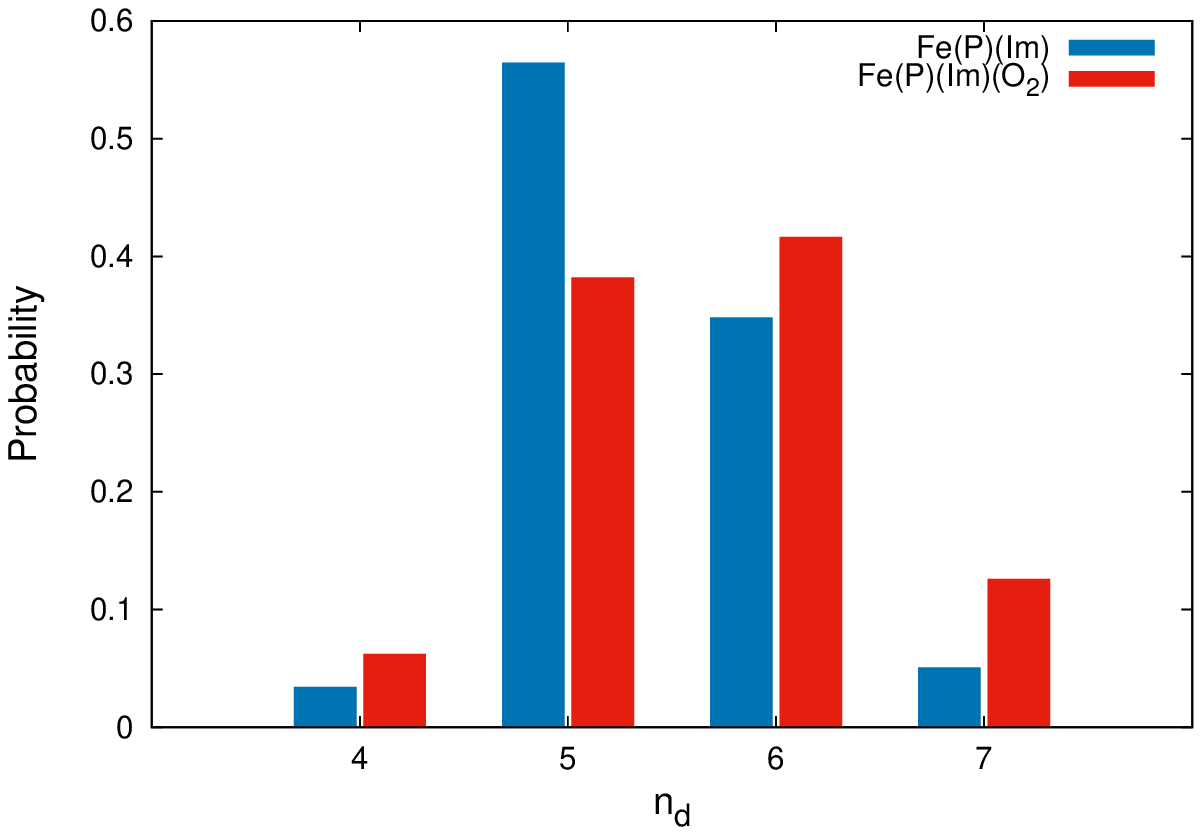}}
\caption{The statistical distribution of the different valence states probabilities for the iron ion for the FeP(Im) and the FeP(Im)(O$_2$) structures.} 
\label{nd}
\end{figure}

The statistical probability of the different atomic configurations of the iron ion obtained by the DMFT calculations for the FeP(Im) and the FeP(Im)(O$_2$) structures is presented in Figure~\ref{nd}.

This result explicitly indicates the presence of the strong valence fluctuations.
As it can be seen from Figure \ref{nd} the centers of mass of the histograms presented for the FeP(Im) and the FeP(Im)(O$_2$) are different.

In the FeP(Im) the dominant contribution to the valence state is provided by $d^5$ configuration (57\%), whereas $d^6$ prevails (42\%) for the FeP(Im)(O$_2$).
It should be noted that for the FeP(Im) the contributions of $d^6$ (35\%) configuration is very significant, as well as for the FeP(Im)(O$_2$) $d^5$ (38\%) and $d^7$ (13\%).
This observation allows one to conclude that the state of the system under consideration has an entangled superposition type and the contribution of the dominant state configuration doesn't exceed~57\%.
Due to the fact that the iron ion $d$-states in this complex porphyrin and ligand surrounding isn't a pure $d$-states, then the observed entangled superposition could be explained as the result of the strong hybridisation~\cite{Scherlis2007a} of the iron states with the states of the neighbouring atoms, as well as by the fact that the possible spin states are energetically very close to each other~\cite{Panchmatia2010a}.
Besides, from the presented figure it can be seen that during the oxidation of the hemoglobin molecule, the iron ion transits to the state with the higher oxidation degree.

\begin{figure}[ht]
\centerline{\includegraphics[width=0.8\linewidth]{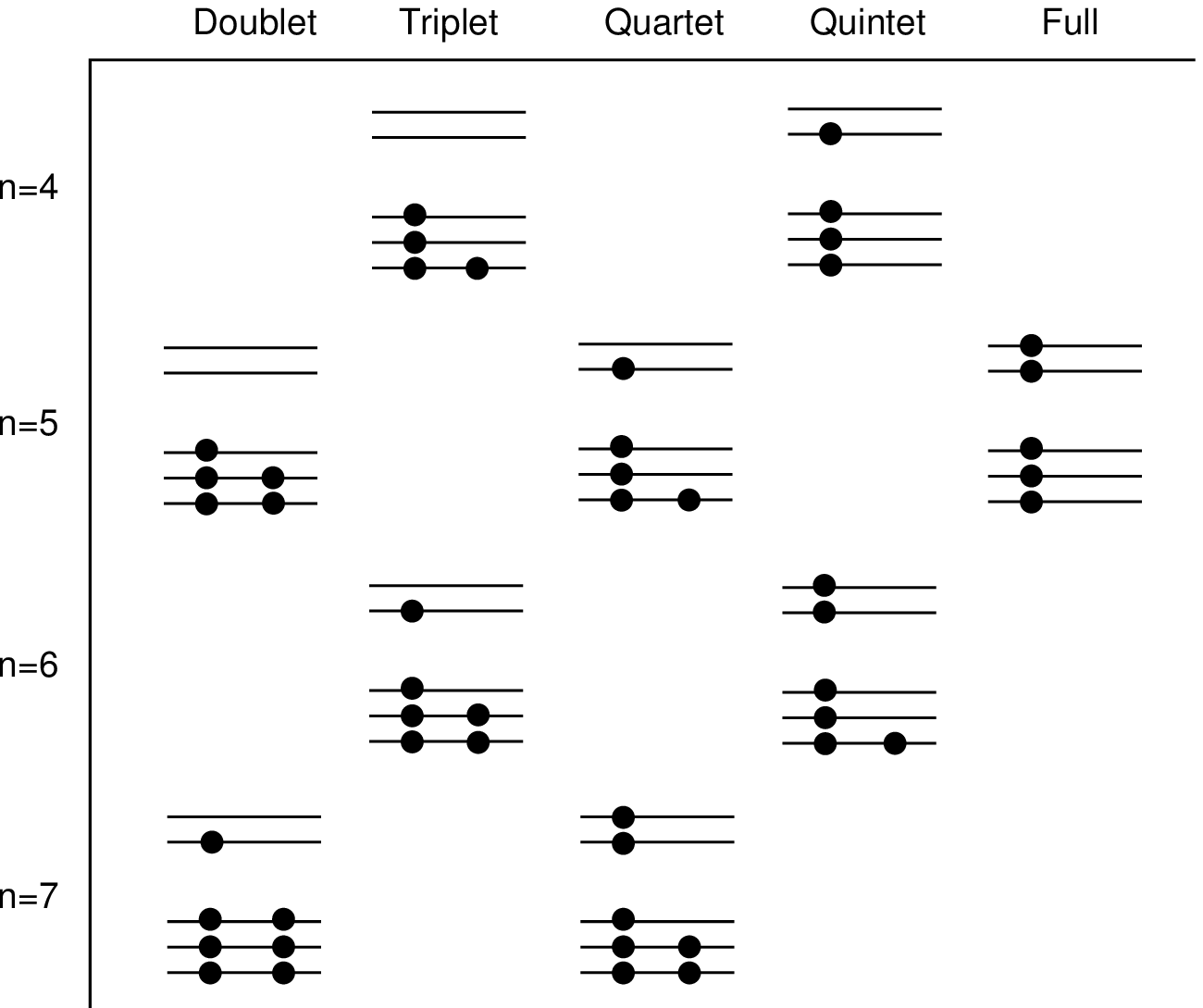}}
\caption{The schematic representation of the different Fe-$d$ electronic configurations.} 
\label{schema}
\end{figure}

\begin{figure}[ht]
\centerline{\includegraphics[width=0.9\linewidth]{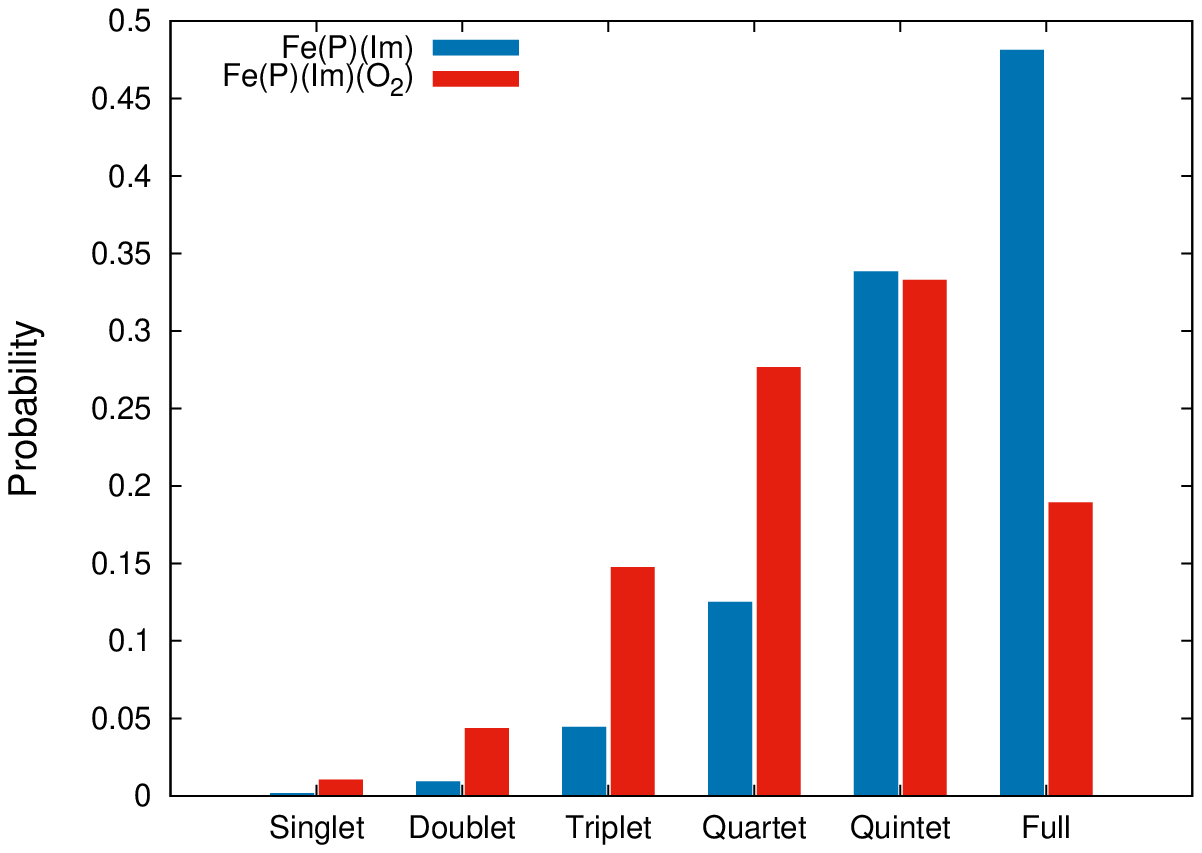}}
\caption{The statistical distribution of the probability for the different spin configurations for the FeP(Im) and the FeP(Im)(O$_2$) systems.} 
\label{sz}
\end{figure}

For every electronic configuration, i.e. for the number of valence electrons in the $d$-shell, several spin configurations are possible. 
Few of them are shown schematically in Figure \ref{schema}. 
We calculated the probability of every spin state for the iron ion in the FeP(Im) and the FeP(Im)(O$_2$) systems.
Resulting histograms are shown in Figure \ref{sz}.

The oxidized and deoxidized HHB molecules have different spin configuration of the Fe ion as it clear from Figure \ref{sz}. The maximum weight has the quintet state in the FeP(Im)(O$_2$) and the fully polarized state in the FeP(Im). At the same time, for both molecules the maximum weight doesn't exceed 48~\%. 

This clearly indicates, that there is the transition between two entangled quantum states during the HHB oxidation process. The spin state of the Fe ion in both, the FeP(Im) and the FeP(Im)(O$_2$) systems, is a mixture of at least four spin configurations with different total moment. The effective spin moment of the Fe ion decreases when oxygen molecule is attached.
The difference of the effective spin moment between the FeP(Im) ($s\approx$2.1) and the FeP(Im)(O$_2$) ($s\approx$1.7) has the value about 0.39.
This result generally is consistent with the data obtained in the work~\cite{Weber2013}. 

\begin{figure}[ht]
\begin{minipage}[h]{0.9\linewidth}
\centerline{\includegraphics[width=\linewidth]{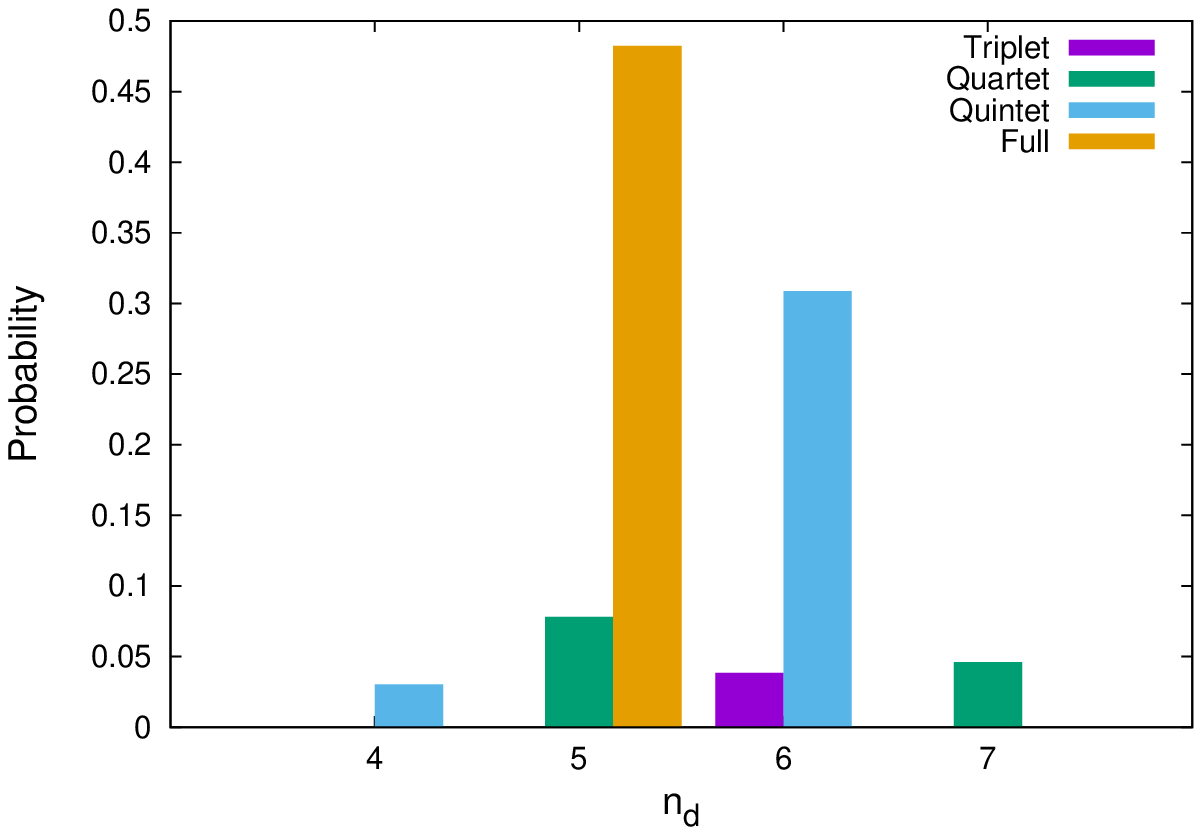}}
\end{minipage}
\vfill
\begin{minipage}[h]{0.9\linewidth}
\centerline{\includegraphics[width=\linewidth]{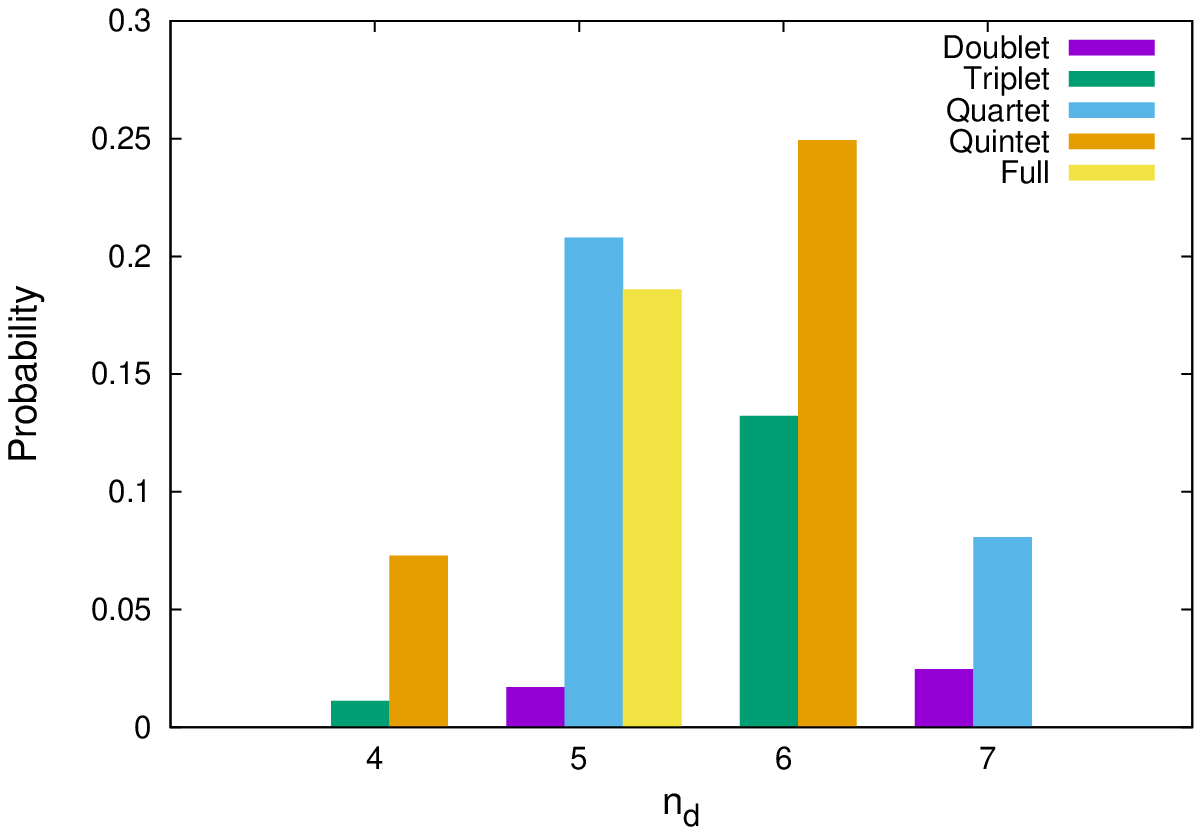}}
\end{minipage}
\caption{The statistical distribution of the probability of the dominant atomic configurations of the iron ion for the FeP(Im) (upper panel) and  the FeP(Im)(O$_2$) (lower panel) systems.} 
\label{ndsz}
\end{figure}

To emphasize the entangled state of the Fe ion, the contributions from the dominant valence and spin configurations to the final state of the ion are shown in Figure \ref{ndsz}.
Here we can see that the $d^5$, and in a less degree the $d^6$, valence configurations contribute to the high-spin state for the FeP(Im) full polarized and the quintet cases respectively. 
Two most probable configurations for the FeP(Im)(O$_2$) are the quintet formed by the $d^6$ and $d^4$ states and the quartet formed by the $d^5$ and the $d^7$ states.

As we can see from Figure \ref{ndsz} the number of the major states with more than 5~\% probability in the FeP(Im) equals three but for the FeP(Im)(O$_2$) it is twice as much that indicates the growth of the quantum entanglement.
This behavior can be caused by the occurrence of the hybridization between the oxygen $p$-states during the molecule oxidation. 
\begin{figure}[ht]
\centerline{\includegraphics[width=0.9\linewidth]{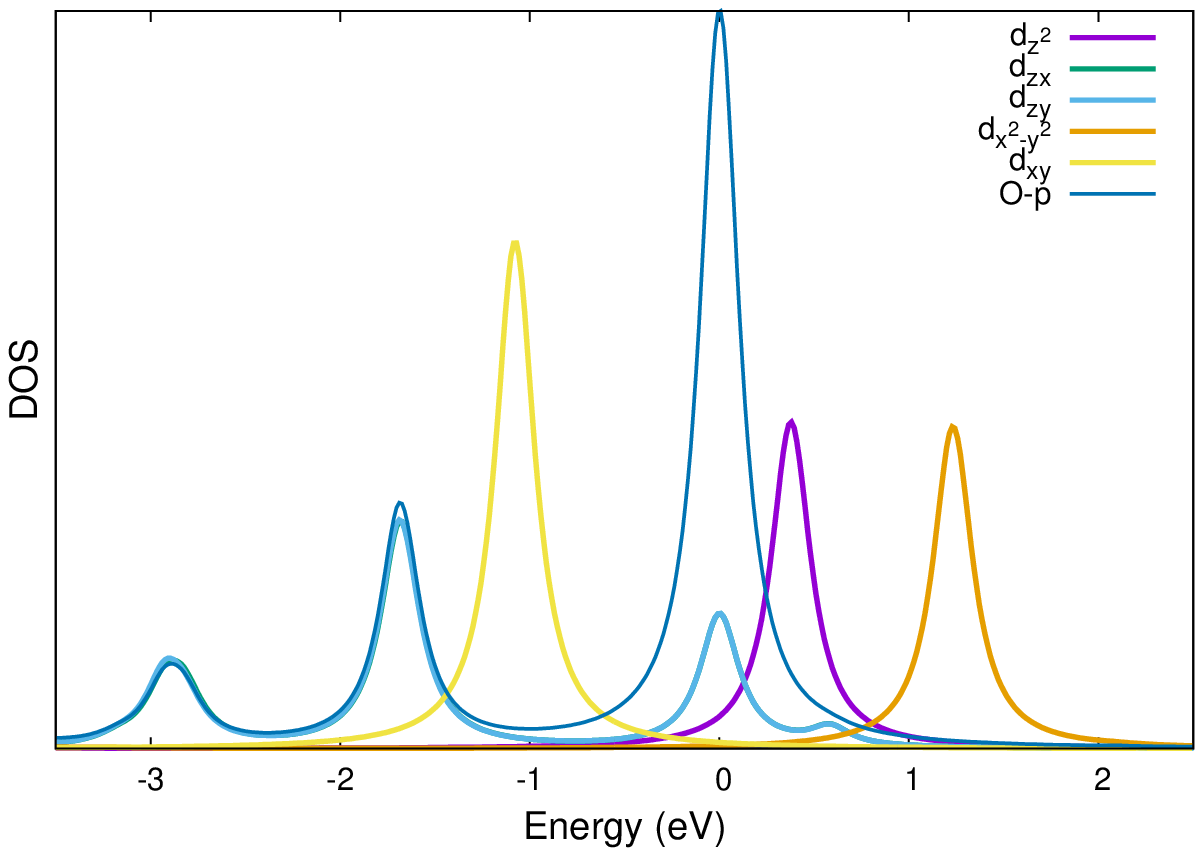}}
\caption{Non-correlated partial density of states of the FeP(Im)(O$_2$) for the Fe-$d$ an O-$p$ states. Fermi level is placed at zero energy.} 
\label{dos}
\end{figure}
Existence of corresponding hybridization effect can be seen in Figure \ref{dos} where the partial density of states for the Fe-$d$ and O-$p$ states is shown in the absence of Coulomb correlations.
From this figure one can see that significant hybridization takes place between the degenerate $d_{zx}$, $d_{zy}$-states of the iron ion and the $p$-states of the oxygen.

\section{Conclusion}

The traditional scenario of the spin state transition accompanying the oxidation process FeP(Im)$\rightarrow$FeP(Im)(O$_2$) from the high-spin state to the low-spin state is simplified and is based on the formal valence value of the iron ion. 
In this work it was shown that the state of the iron atom placed in the active center of the hemoglobin molecule is entangled. 
In the both systems, oxidized and deoxidized, the Fe ion state is the superposition of states with various d-shell occupations and spin moment values.
One can observe the local valence fluctuations of the metal ion.
The spin transition during the formation the bond with the oxygen occurs between two entangled electronic states of the Fe ion.  
Therefore the oxidation of the FeP complex results in a tuning of a delicate balance between several possible electronic and magnetic configurations of the active center. 
The process is accompanied by the decrease of the effective spin moment.

\begin{acknowledgement}
This work was supported by the Russian Foundation for Basic Research (RFBR) for the project No. 14-02-31325, within the state assignment of FASO of Russia (theme <<Electron>> No. 01201463326) as well as the Russian Science Foundation grant project No. 14-22-00004. Calculations were performed using Uran supercomputer of IMM UB RAS.

The authors are grateful to A.I.~Poteryaev and to A.A.~Dyachenko for valuable discussions.
\end{acknowledgement}


\bibliography{achemso}

\end{document}